\documentclass[useAMS,usenatbib,usegraphicx]{mn2e}
\usepackage{times}

\title{The origin of the lead-rich stars in Galactic halo: investigation of the model parameters for the s-process}
\author[Wenyuan Cui, and Bo Zhang]{Wenyuan Cui$^{1,2}$\thanks{E-mail:
wenyuancui@126.com.cn} and Bo
Zhang$^{1,2}$\thanks{E-mail: zhangbo@mail.hebtu.edu.cn}\\
$^{1}$National Astronomical observatories, Chinese Academy of Sciences, Beijing 100012, China\\
$^{2}$Department of Physics, Hebei Normal University, 113 Yuhua Road,
Shijiazhuang 050016, China}
\begin{document}

\date{Received ........; accepted ........}

\pagerange{\pageref{firstpage}--\pageref{lastpage}} \pubyear{2005}

\maketitle

\label{firstpage}

\begin{abstract}
Several stars at the low-metallicity extreme of the Galactic halo
show large spreads of [Pb/hs]. Theoretically, a s-process pattern
should be obtained from an AGB star with fixed metallicity and
initial mass. For the third dredge-up and the s-process model,
several important properties depend primarily on the core mass of
AGB stars. \citet{b40} reported that the initial-final-mass
relation steepens at low metallicity, due to low mass-loss
efficiency. This perhaps affects the model parameters of the AGB
stars, e.g. the overlap factor and the neutron irradiation time,
in particular at low metallicity. The calculated results show
indeed that the overlap factor and the neutron irradiation time
are significantly small at low metallicities, especially for
3.0M$_{\odot}$ AGB stars. The scatter of [Pb/hs] found in low
metallicities can therefore be explained naturally when varying
the initial mass of the low-mass AGB stars.
\end{abstract}


\section{Introduction}

The elements heavier than the iron peak are made through neutron
capture via two principal processes: the r-process (for rapid
process) and the s-process (for slow neutron capture
process)\citep{b6}. The observations have confirmed that, indeed,
Asymptotic Giant Branch (AGB) stars show overabundances of
elements heavier than iron at their surface \citep{b34}, which
clearly indicate that the s-process takes place during the AGB
phase in the evolution of low- and intermediate-mass stars
(0.8$\leq$M(M$_{\odot}$)$\leq$8). Low-mass AGB stars are usually
thought as the main nuclear production site of the s-process
elements \citep{b13,b31,b19}. By now, the generally favoured
s-process model is associated with the partial mixing of protons
(PMP) into the radiative C-rich layers during thermal pulses
\citep{b35,b13,b15,b36}. PMP activates the chain of reactions
$^{12}$C(p,$\gamma$)$^{13}$N($\beta$)$^{13}$C($\alpha$,n)$^{16}$O
which likely occurs in a narrow mass region of the He intershell
(i.e. $^{13}$C-pocket) during the interpulse phases of an AGB
star. The s-elements thus produced in the deep interior by
successive neutron captures are subsequently brought to the
surface by the third dredge-up. Using the primary-like neutron
source ($^{13}$C($\alpha$,n)$^{16}$O) and starting with a very low
initial metallicity, most iron seeds are converted into
$^{208}$Pb. When the third dredge-up episodes mix the neutron
capture products into the envelope, the star will appear
s-enhanced and lead-rich. If the standard PMP scenario holds, all
s-process-enriched AGB stars with metallicities
[Fe/H]\footnote{where [X/Y]$=$log(N$_{X}$/ N$_{Y}$) - log(N$_{X}$/
N$_{Y}$)$_{\odot}$, $\odot$ refers to the Solar System
abundances.}$\leq$ -1.3 are thus predicted to be lead(Pb) stars
([Pb/hs]$\geq$ 1, where hs denotes the 'heavy' s-process elements
such as Ba, La, Ce), independently of their initial mass and
metallicity \citep{b13,b17,b18}.

The first three such lead stars (HD187861, HD224959, HD196944 ),
have been later confirmed by \citet{b38}. At the same time,
\citet{b1} found that the slightly more metal-deficient stars LP
625-44 and LP 706-7 are enriched in s-elements, but cannot be
considered as lead stars ([Pb/Ce]$<$ 0.4), in disagreement with
the standard PMP predictions. Recently, more spectroscopic data of
s-rich and lead-rich stars are reported
\citep{b2,b11,b30,b39,b25,b26,b33}. The large observation data
spreads of [Pb/hs] are strong indication to suspect a large
intrinsic spread of integrated neutron irradiations. In order to
explain the spreads of [Pb/hs], a large spread of $^{13}$C-pocket
efficiencies is subsequently proposed by \citet{b36}. However, it
should be stressed here that the predictions of the standard PMP
scenario are rather robust \citep{b17}. In the framework of the
PMP scenario, there is no obvious degree of freedom that could be
used to reduce the lead production in low-metallicity AGB stars
\citep{b39}. At present, the physical explanation for the
different $^{13}$C-pocket strengths, which perhaps should not be
consistent with the primary nature of the  neutron source, is not
yet found \citep{b32}. Thus the fundamental problems, such as the
formation and the consistency of the $^{13}$C-pocket, the neutron
exposure signature in the interpulse period, currently exist in
the models of AGB stars.

For the third dredge-up and the thermal pulse model, several
important properties depend primarily on the core mass M$_{c}$,
while the dependence on the other stellar parameters is negligible
or marginal \citep{b22,b16,b28}. \citet{b40} reported that the
initial-final-mass relation steepens at low metallicity, due to
low mass-loss efficiency. This may cause the degenerate cores of
low-\textit{Z}, high-mass AGB stars to reach the Chandresekhar
mass, leading to an Iben \& Renzini-type-1.5 supernova
\citep{b23}. On the other hand, this can obviously affect model
parameters of the AGB stars, e.g. the overlap factor \textit{r},
which is the fraction of material that remains to experience
subsequent neutron exposures, and the neutron irradiation time
$\Delta t$, in particular at low metallicity.

In this paper we will present a calculation which indicates that
at low metallicity, large core mass of AGB stars may allow the
overlap factor and the duration of neutron irradiation to reach
small values for a fixed initial mass of AGB stars. This will
affect the characters of the s-process nucleosynthesis, and can
explain the observed abundance pattern of lead-rich stars. The
next section discusses the model parameters of AGB stars. In
section 3, we discuss the characteristics of the s-process at low
metallicity and the possibility of lead stars. Finally, in section
4 we summarize the main conclusion that can be drawn from such an
analysis.

\section[]{Model Parameters of AGB Stars}

There are four parameters in the parametric model of \citet{b21}
on s-process nucleosynthesis. They are the neutron irradiation
time $\Delta$t, the neutron number density \textit{N$_{n}$}, the
temperature \textit{T$_{9}$}(in units of 10$^{9}$ K) at the onset
of the s-process, and the overlap factor \textit{r}. Combining
these quantities, we can obtain the neutron exposure,
\textit{$\Delta$$\tau$=N$_{n}$v$_{T}$$\Delta$t}, where
\textit{v$_{T}$} is the average thermal velocity of neutrons at
\textit{T$_{9}$}. The temperature is fixed at a reasonable value
for the $^{13}$C($\alpha$,n)$^{16}$O reaction,
\textit{T$_{9}$}=0.1, for these studies. Using the initial-final
mass relations as function of metallicity presented by
\citet{b40}, the effects on the parameters can be derived.

\subsection{The overlap factor}

\begin{figure}
% \vspace{59pt}
 \centering
 \includegraphics[width=0.5\textwidth,height=0.25\textheight]{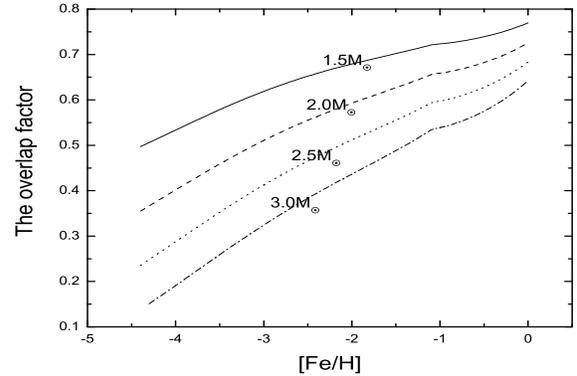}
 \suppressfloats[t]
 \caption{The overlap factor of different initial mass AGB stars, as function of metallicity.}
 %\label{appenfig}

\end{figure}

For the AGB model, the overlap factor \textit{r} is a fundamental
parameter. An analytical formula was given by \citet{b22} as a
function of the core mass M$_{c}$ in the range
0.6$\leq$M$_{c}$$\leq$1.36:
\begin{equation}
   r=0.43-0.795(M_{c}-0.96)+0.346(M_{c}-0.96)^{2}.
\end{equation}
We can obtain an initial-final mass relation as a function of
metallicity, by fitting results of \citet{b40}:
\begin{equation}
   M_{c}=A(Z)+B(Z)M,
\end{equation}
where
\begin{equation}
   A(Z)=0.46449+0.03279log\frac{Z}{Z_{\odot}}+0.00044(log\frac{Z}{Z_{\odot}})^{2},
\end{equation}
and
\begin{equation}
   B(Z)=0.08729-0.04851log\frac{Z}{Z_{\odot}}+0.00468(log\frac{Z}{Z_{\odot}})^{2},
\end{equation}
which is valid for -4$\leq$$log\frac{Z}{Z_{\odot}}$$\leq$0.
Combing the equations (1), (2), (3) and (4), we obtain the overlap
factor as a function of the initial mass and metallicity. The
overlap factor is shown in Fig. 1, which is significantly small at
low metallicities, especially for 3M$_{\odot}$ AGB stars. In AGB
stars with initial mass in the range M = 1.5 $\sim$
3.0M$_{\odot}$, the core mass M$_{c}$ lies between 0.6 and
1.4M$_{\odot}$ at [Fe/H]= -2.5. According to the equation (1), the
corresponding values of \textit{r} will range between 0.8 and
0.13. \citet{b13} have found an overlap factor of \textit{r} = 0.4
$\sim$ 0.7 in their standard evolution model of low-mass AGB stars
at solar metallicity, which lies in our predicted range of r.
\citet{b1} have reported an overlap factor of $r$ $\sim$ 0.1,
found for the best fit to metal-deficient AGB stars that produced
the abundance patterns of LP 625-44 and LP 706-7. In an evolution
model of AGB stars, a small $r$ may be realized if the third
dredge-up is deep enough for s-processed material to be diluted by
extensive admixture of unprocessed material. \citet{b28} have
found that the third dredge-up is more efficient for the AGB stars
with larger core mass. Taking account of the core-mass dependence,
the wide range of $r$-values of the lead enhanced stars can be
explained naturally by the wide range of core-mass values of AGB
stars at low metallicity.

\subsection{The neutron exposure}

\begin{figure*}
 \centering
 \includegraphics[width=0.45\textwidth,height=0.25\textheight]{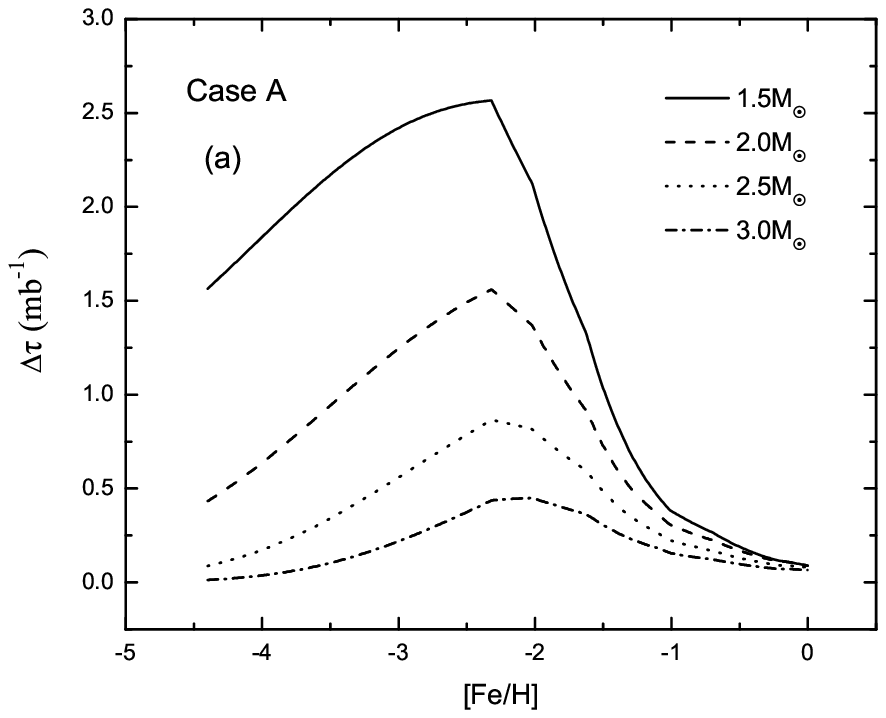}
 \includegraphics[width=0.45\textwidth,height=0.25\textheight]{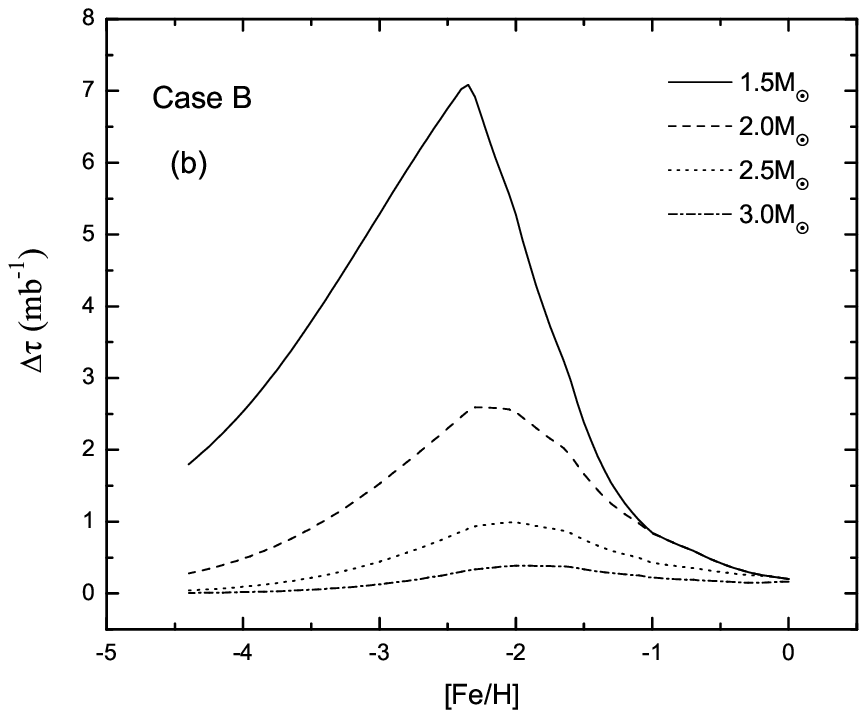}
 %\suppressfloats[t]
 \caption{The neutron exposure of different initial mass AGB stars, as function of metallicity.}
 %\label{appenfig}

\end{figure*}

\citet{b13} have pointed out that the neutron density is
relatively low, reaching $\sim$10$^{7}$ cm$^{-3}$ at solar
metallicity. Since the $^{13}$C neutron source is of primary
nature, the typical neutron density in the nucleosynthesis zone
scales roughly as $1/Z^{0.6}$, from Z$_{\odot}$ down to
$1/50$Z$_{\odot} $. At lower metallicities, the effect of the
primary poisons prevails \citep{b14,b7}.

There is a possibility for the synthesis of s-process elements in
the AGB stars, i.e., with nucleosynthesis taking place during
thermal pulses \citep{b1}. In this case, the neutron irradiation
is derived primarily by the $^{13}$C($\alpha$,n)$^{16}$O reaction,
with a minor contribution from the marginal burning of $^{22}$Ne.

However, in the s-process scenario that invokes radiative
$^{13}$C-burning, the nucleosynthesis mostly occurs during the
relatively long interpulse period, in a thin radiative layer at
the top of the He intershell (i.e., the $^{13}$C-pocket model). A
second neutron burst giving rise to a small neutron exposure is
released by the marginal activation of the $^{22}$Ne neutron
source in the convective thermal pulse. The neutron irradiation
time \textit{$\Delta$t} should be close to the interpulse period
at low metallicity due to the combination of two reasons. The
first is that the higher neutron density can lead to longer
neutron irradiation time, and the second is that the shorter
interpulse period is expected for larger core-mass of AGB stars
\citep{b16}. Therefor, adopting the interpulse period as the
neutron irradiation time will have a smaller effect on the
low-metallicity stars of interest here than that on stars of solar
metallicity.

In our calculation, the neutron irradiation time $\Delta t$ is
adopted respectively as follows:\\
Case A: the duration of the thermal pulse, where the
core-mass-duration of the convective shell relation is adopted
from \citet{b22}.\\
Case B: the interpulse period, where the core-mass-interpulse period
relation is adopted from \citet{b5}.\\
We choose respectively $\Delta\tau$=0.08mb$^{-1}$ at [Fe/H]= -0.3
in case A, which corresponds to a mean neutron exposure
$\tau_{0}$=0.296(\textit{T$_{9}$}/0.348)$^{1/2}$mb$^{-1}$, and
$\Delta\tau$=0.2mb$^{-1}$for 3.0M$_{\odot}$ AGB stars with solar
metallicity in case B \citep{b13}. Using the initial-final mass
relations given by \citet{b40} and the neutron irradiation time
$\Delta t$, we can obtain the neutron exposure $\Delta\tau$ as a
function of metallicity and initial mass (see Fig.2). The trend
shown in Fig.2 (Case A and Case B) can be understood as follows:
$\Delta\tau$ is proportional to the neutron number density
\textit{N$_{n}$} and the neutron irradiation time
\textit{$\Delta$t}, where \textit{N$_{n}$} is expected to increase
with declining metallicity. However, \textit{$\Delta$t} declines
with declining metallicity due to the increasing of stellar core
mass, which directly leads to a decline of $\Delta\tau$ at very
low metallicity, especially for 3M$_{\odot}$ AGB stars.

Based on the primary nature of the $^{13}$C neutron source, the
value of $\Delta\tau$ will reach about 6.3 mb$^{-1}$ around
[Fe/H]$\simeq$ -2.5 for the case of radiative  $^{13}$C-burning
\citep{b14}. Our result of case B for the 1.5M$_{\odot}$ AGB stars
 is close to the above value. Because the neutron irradiation
time $\Delta t$ is shorter for the larger AGB stars, the neutron
exposure $\Delta\tau$ should be smaller too. \citet{b1} have
reported a neutron exposure, $\Delta\tau\sim$ 0.7 mb$^{-1}$ for
metal-deficient stars LP 625-44 and LP 706-7, which is in the
range of our calculated results for the both cases. The results
shown in Fig. 2 imply that the wide range of $\Delta\tau$ can be
obtained naturally by considering the dependence of the
irradiation time on the core mass . Since the Pb abundance is very
sensitive to the neutron exposure \citep{b15,b31}, large
variations of the [Pb/hs] ratio could be expected.

\section{Discussion}

\subsection{The Case A}

\begin{figure*}
 \centering
 \includegraphics[width=0.45\textwidth,height=0.25\textheight]{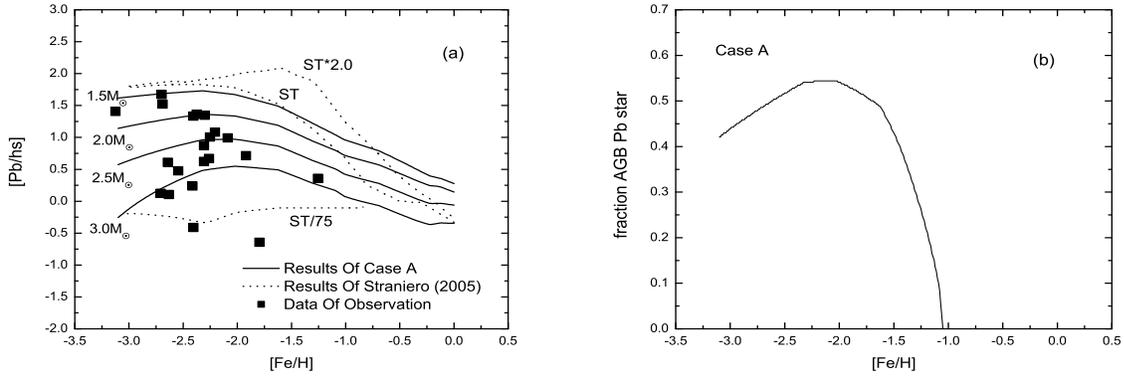}
 \includegraphics[width=0.45\textwidth,height=0.25\textheight]{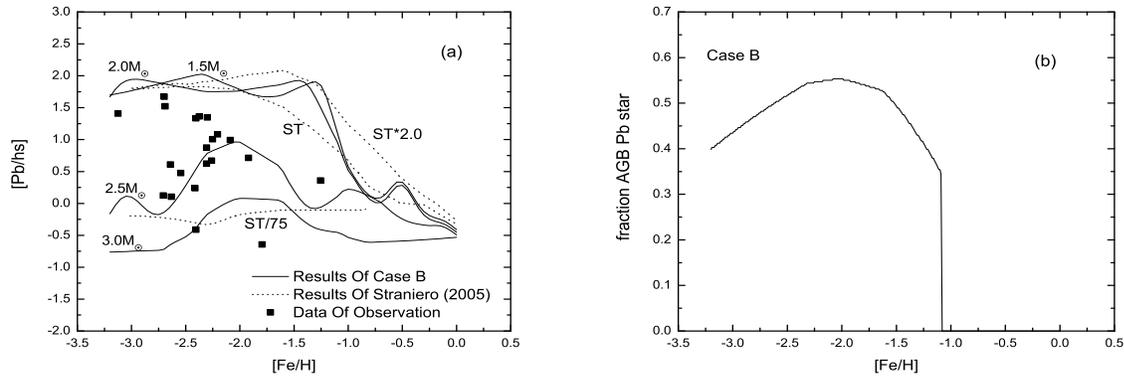}
 \suppressfloats[h]
 \caption{(a) Predicted [Pb/hs] versus [Fe/H] for different initial stellar mass choices (solid
lines). Plot lines represent the results predicted by \citet{b36}
for their ST, ST$\times$2 and ST/75 case respectively.
Spectroscopic data of s-rich and lead-rich stars adopted from
\citet{b12} are also included for comparison. (b) The ratio of
lead stars over the low mass AGB stars (1.5-3.0M$_{\odot}$), as
function of metallicity.}

\end{figure*}

\begin{figure*}
 \centering
 \includegraphics[width=0.45\textwidth,height=0.25\textheight]{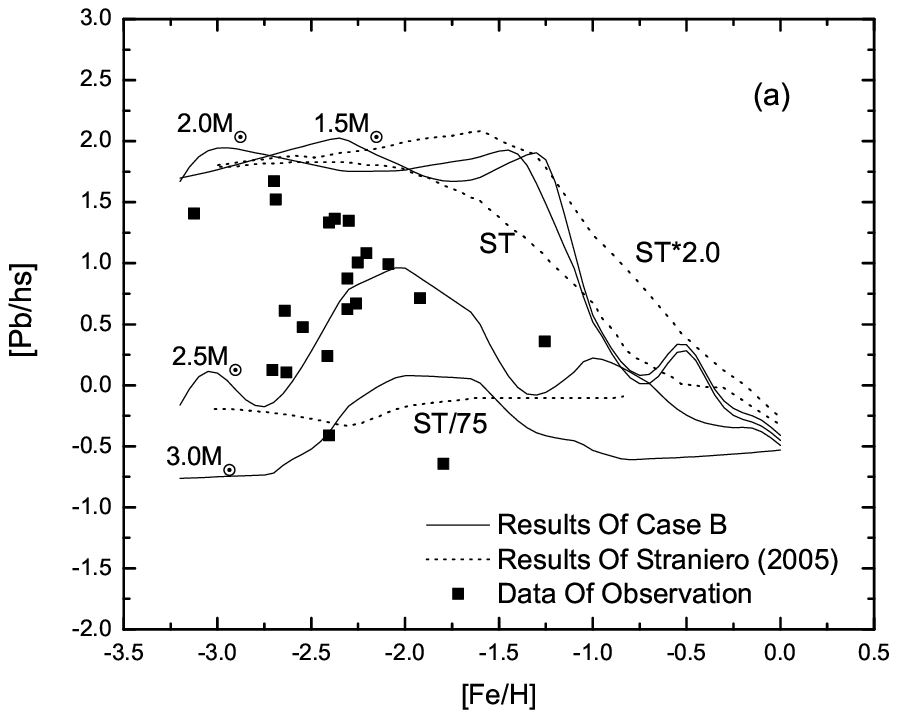}
 \includegraphics[width=0.45\textwidth,height=0.25\textheight]{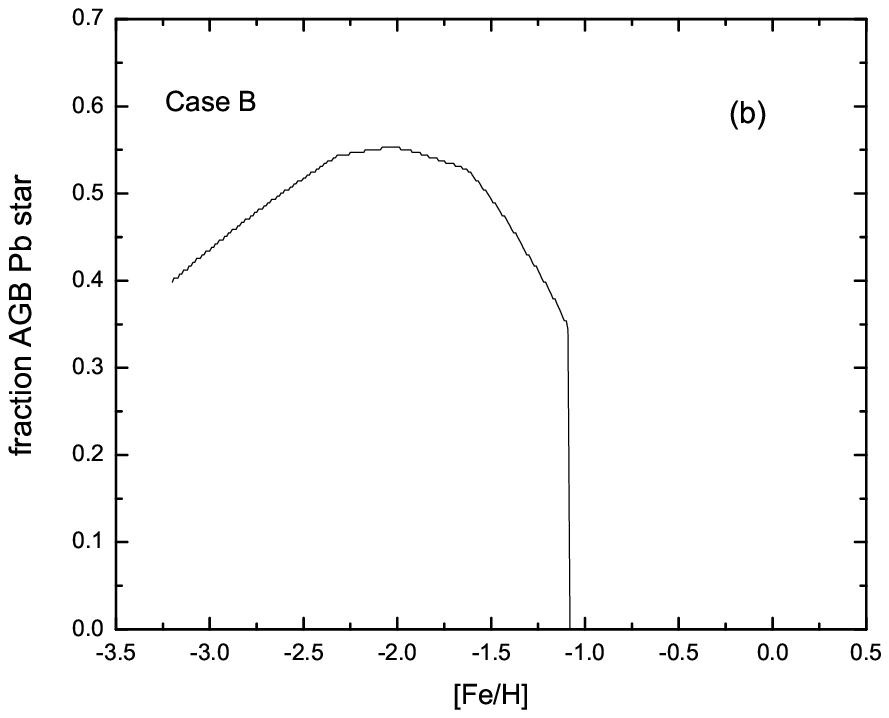}
 %\suppressfloats[b]
 \caption{(a) Predicted [Pb/hs] versus [Fe/H] for different initial stellar mass choices (solid
lines). Plot lines represent the results predicted by \citet{b36}
for their ST, ST$\times$2 and ST/75 case respectively.
Spectroscopic data of s-rich and lead-rich stars adopted from
\citet{b12} are also included for comparison. (b) The ratio of
lead stars over the low mass AGB stars (1.5-3.0M$_{\odot}$), as
function of metallicity.}

\end{figure*}

The first model for $^{13}$C-burning in AGB stars assumed that the
neutrons were released in convective conditions \citep{b20,b27}.
In such calculations, a repeated neutron exposure was achieved
thanks to partial overlapping of material cycled through several
thermal pulses. The s-process mechanism could be approximated by
an exponential distribution of neutron exposures $\propto$
exp($-\tau/\tau_{0}$), where the mean neutron exposure is given by
$\tau_{0}$= $-\Delta\tau/ln\textit{r}$. The final abundance
distributions depend mainly upon $\tau_{0}$.

In order to investigate the efficiency of the s-process site,
[Pb/hs] is particularly useful \citep{b36}. There have been many
theoretical studies of s-process nucleosynthesis in low-mass AGB
stars \citep{b12,b24,b36}. Unfortunately, the precise mechanism
for chemical mixing of protons from the hydrogen-rich envelop into
the $^{12}$C-rich layer to form  $^{13}$C-pocket is still unknown
\citep{b32}. This makes it even harder to understand a large
spread of [Pb/hs] found in carbon-rich, metal-deficient stars. It
is an interesting exercise to investigate the effect of the
parameters presented above upon the s-process efficiency of AGB
stars. For this purpose, we have used the simple analytical
formulation \citep{b9,b10} without depending on any specific
stellar model, with many of the neutron-capture rates updated
\citep{b4}, to study what physical conditions are possible to
reproduce the observed abundance pattern found in the metal-poor
stars. The variation of the logarithmic ratio [Pb/hs] with
metallicity is shown in Fig. 3a, where solid lines represent
respectively results of different initial mass of AGB stars. As a
comparison, spectroscopic measurements (filled squares) of C and
s-rich metal-poor stars are reported. Because of the uncertainties
related to the formation mechanism of the $^{13}$C-pocket
\citep{b8}, a large spread of $^{13}$C-pocket efficiencies has
been proposed by \citet{b36} in order to explain the spreads of
[Pb/hs], which has proved to be effective for several purposes
\citep{b13,b37,b8}. The results (plot lines) predicted by
\citet{b36} for their standard case (hereafter ST), ST$\times$2
and ST/75 case are also presented respectively. The ST case
\citep{b13} was shown to reproduce remarkably well the abundance
distribution of the solar main component at metallicity
Z=Z$_{\odot}$/2 \citep{b3}.

Assuming a Salpeter IMF with $\alpha$= -2.35, the ratio of lead
stars compared to the low mass AGB stars' (1.5-3.0M$_{\odot}$) is
shown in Fig. 3b.

\subsection{The Case B}

We use the parametric approach based on the model of low-mass AGB
stars computed by \citet{b13} assuming that all the pulses are
identical. In this case,  the neutron exposure is not well
approximated with the exponential distribution, and the final
s-process abundance distributions depend mainly upon the neutron
exposure $\Delta\tau$, the mass fraction of $^{13}$C-pocket in the
He intershell $q$ (adopted as 0.05) and overlap factor $r$. The
adopted initial abundances of seed nuclei lighter than the iron
peak elements were taken to be the solar-system abundances, scaled
to the value of [Fe/H]. Because the neutron-capture-element
component of the interstellar gas that formed very metal-deficient
stars is expected to consist of mostly pure r-process elements,
for the other heavier nuclei we use the r-process abundances of
the solar system \citep{b3}, normalized to the value of [Fe/H]. We
carry out s-process nucleosynthesis calculation by means of an
extensive reaction network described earlier \citep{b29}. The
results are shown in Fig. 4.

It results that for very metal-poor stars, large spreads of
[Pb/hs] are predicted for both cases. The agreement of the results
with the observations provides a strong support to the validity of
the parameters adopted in this work. Here, the logarithmic ratio
[Pb/hs] shows a complex trend versus [Fe/H], due to the conjunct
effect of the overlap factor and neutron exposure. Although the
parameters of the two cases are different in part, the ratios of
lead stars obtained in this work are very similar. At [Fe/H]
$\simeq$ -2.0, the ratios of both cases all reach the maximum
value around 0.55. The results obtained in this paper are the
evidence that maybe the well established theories of the s-process
nucleosynthesis, as it works in solar-like metallicities, may not
work well at extremely low metallicity because of greatly changed
parameters. In fact, such results could be mainly led to by the
new initial-final mass relations \citep{b40}, i.e. the inefficient
mass loss of the AGB stars at low metallicity.

\section{Conclusion}

Theoretically, a s-process pattern should be obtained from an AGB
star with fixed metallicity and initial mass. Taking account of
the core-mass dependence, the large intrinsic spread of integrated
neutron irradiations for the AGB stars at low metallicities is
obtained, then the scatter of [Pb/hs] such as found in low
metallicities can therefore be explained naturally when varying
the initial mass of the AGB stars. Based on the relation of the
overlap factor with the initial mass of the AGB stars, we can
speculate that the Pb stars are polluted by low mass AGB stars
(e.g. 1.5-2.5M$_{\odot}$) and the non-Pb stars ([Pb/hs]$<$1) are
polluted by larger mass AGB stars. We remind the reader that
though the parametric approach is still useful to interpret
observation data, it does not refer to detailed stellar evolution
models. Obviously, a more precise overlap factor-core mass law and
a more precise neutron irradiation time-core mass relation still
have to wait for new models of nucleosynthesis in AGB stars.

\section*{Acknowledgments}

We are grateful to the referee for the very valuable comments and
suggestions which improved this paper. This work is supported by
the National Natural Science Foundation of China under grant No.
10373005.

\bsp

\label{lastpage}

\end{document}